# Research on accurate stereo portrait generation algorithm of scientific research team

Mingying Xu, Junping Du, Meiyu Liang, Zhe Xue, Ang Li

(*Beijing University of Posts and Telecommunications, Beijing Key Laboratory of Intelligent Telecommunication Software and Multimedia, Beijing, China 100876*)

**Abstract** In order to smoothly promote the establishment of scientific research projects, accurately identify the excellent scientific research team, and intuitively and comprehensively describe the scientific research team, it is of great significance for the scientific research management department to comprehensively understand and objectively evaluate the scientific research team. At present, the research work on the construction of accurate three-dimensional portrait of scientific research team is relatively less. In view of the practical demand of scientific research management department, this paper proposes an accurate stereo portrait generation algorithm of scientific research team. The algorithm includes three modules: research team identification, research topic extraction and research team portrait generation. Firstly, the leader of the scientific research team is identified based on the iterative middle centrality ranking method, and the members of the scientific research team are identified through the 2-faction and snowball methods, so as to realize the identification of the scientific research team. Then, considering the statistical information of words and the co-occurrence features of words in the research team, the research topics of the research team are extracted to improve the accuracy of research topic extraction. Finally, the research team portrait generation module generates the accurate three-dimensional portrait of the research team through the generation of the research team profile, the construction of the research cooperation relationship, and the construction of the research team topic cloud. The research team is identified on the data set of scientific research achievements, and the accurate three-dimensional portraits of the research team are generated and visualized. Experiments verify the effectiveness of the proposed algorithm.

## 1 Introduction

Teamwork has become imperative in many fields. Over 90% of publications in science, technology and engineering are finished collaboratively. With the increasing complexity and specialization of scientific research tasks[1], scientific research tasks are more and more inclined to be finished by scientific research teams. Scientific research team cooperation helps team members achieve information resource sharing among research teams, acquire and integrate knowledge, skills, materials and other necessary resources[2,3] to deal with scientific problems, which also improves research productivity and improves the output rate of scientific research results. It is also the key to enhance scientific and technological innovation capabilities and forge successful career paths for researchers. For example, Ding et al. found that highly productive scholars in the field of information retrieval are more willing to collaborate and are more willing to collaborate with people with similar research interests[4], and related surveys show that teamwork has been increasing over the past century[5].

In order to set up major projects, the scientific research management department needs to find excellent scientific research teams. With the rapid development of science and technology, a large amount of scientific and technological big data has also appeared in scientific research, and it has the characteristics of complexity, structure, multi-source, and heterogeneity[6]. How to quickly and accurately identify scientific research teams from the vast sea of scientific and technological big data and how to effectively describe the scientific research team so that the team information can be displayed more intuitively, and how to provide scientific decision support for scientific research team management and review expert decision-making are important scientific issues that need to be solved urgently.

At present, the scientific research management department mainly uses the application form submitted by the scientific research team and the listed academic achievements as the basis for the identification of the research team and the evaluation of the strength of the

This work is supported by National Key R&D Program of China (2018YFB1402600)
**Corresponding Author**：Junping Du（junpingdu@126.com）

team. Effectively identifying scientific research teams from massive scientific and technological resources, analyzing and describing[7] them will help to quickly grasp the current situation of scientific research teams, fully understand the teams, and then improve the management of scientific research teams and better guide the development of scientific research teams[8].

This paper proposes an accurate stereo portrait generation algorithm for scientific research teams. The algorithm identifies scientific research teams by acquiring scientific research results and other related scientific and technological resources based on the idea of social network analysis[9], and uses iterative intermediate centrality ranking method to identify team leaders and team member. According to the characteristics of the research team's achievement data[10,11,12], the research topics of the research team are extracted by combining the statistical characteristics and co-occurrence characteristics[13,14] of words. Combine generated introduction of the scientific research team, the cooperation relationship diagram of the scientific research team and the word cloud of the research topic of the scientific research team, finally generate the accurate stereo portrait of the scientific research team, which comprehensively and accurately described the scientific research team and provide reference for other relevant departments.

The main contributions of this paper are as follows:

1) Accurate stereo portrait algorithm of scientific research teams is proposed, and the portrait of scientific research teams are constructed through three modules: identification of scientific research teams, extraction of research topics of scientific research teams, and generation of accurate stereoscopic portraits of scientific research teams;

2) An unsupervised topic extraction method for academic paper data is proposed, which comprehensively considers the statistical information of words and the semantic information between words;

3) Experiments were carried out on the real scientific research results data set, and the relevant scientific research teams were accurately excavated and the research topics of the scientific research team were extracted, and the template was designed to generate an accurate stereo portrait of the scientific research team.

## 2 Related Works

In recent years, people use large-scale scientific research literature databases to establish a scientific research cooperation network to identify scientific research teams has been widely studied. It is an important method to identify scientific research teams by constructing an academic cooperation network[15,16] through academic cooperation, and then identifying scientific research teams through social network analysis[17,18]. This type of method usually uses information such as author co-authorship and literature citations to construct an overall scientific research cooperation network[19]. A large part of the research is mainly based on static analysis methods to analyze the statistical characteristics of the scientific research cooperation network and the personal attributes of the cooperative members[20]. Choose betweenness centrality as an indicator to identify team leaders, and dig out the team leaders and team members who cooperate closely in the scientific research network through the ranking of betweenness centrality indicators, and visualize the identification results. The ability to judge and enhance the credibility of team identification results has been widely used in the identification of scientific research team leaders, and has achieved good results. The above method has the problem that one team is mistakenly identified as multiple teams, which affects the identification effect of the scientific research team. In order to solve this problem, based on the iterative intermediate centrality method, the leader of the scientific research team is identified according to the heuristic rules and the structural characteristics of the scientific research cooperation network, and the members of the scientific research team are identified by the 2-faction and the snowball method, so as to realize the identification and mining of the scientific research team more accurately[21].

Analyzing and characterizing research teams can help improve the management of research teams. The extraction of the research topics of the scientific research team is an important part of describing the overall overview of the scientific research team. At this stage, topic extraction methods mainly include unsupervised methods and supervised methods[22]. Unsupervised topic extraction methods do not require labeled datasets[23] that have been labeled with topics, but exploit the underlying structure of text documents, usually using statistics[24] or graph-based methods[25] to extract the keywords. Supervised keyword extraction methods typically train models using machine learning or deep learning techniques[26] on labeled datasets. Supervised keyword extraction algorithms require high labor costs, and unsupervised methods do not require human-annotated corpora, which are convenient and fast, and thus have been widely used[27].

The method of describing the overall characteristics of the scientific research team has not been widely studied. With the development of user portrait technology[28,29], based on multiple heterogeneous data sources such as personal homepage and howNet, the method of constructing a portrait of a scientific researchers has been proposed and has been extensively studied[30,31]. On this basis, the concept of a portrait of a scientific research team has also been proposed[32]. By integrating multi-source heterogeneous data, the construction includes the extraction of team structure, team closeness extraction, team research topic extraction[33], and team influence extraction of scientific research team portraits. The existing research team portrait construction method is to collect the existing research team data from personal homepage, Baidu Encyclopedia, etc. to construct the scientific research team portrait. This method is time-consuming and laborious, and the obtained team information may be incomplete, thus affecting the effect of portrait construction.

## 3 Proposal of accurate stereo portrait algorithm for scientific research team

This paper proposes an algorithm for generating accurate stereoscopic portraits of scientific research teams. The overall framework is shown in Figure 1, which includes three modules: identification of scientific research teams, extraction of research topics of scientific research teams, and generation of accurate stereoscopic portraits of scientific research teams.

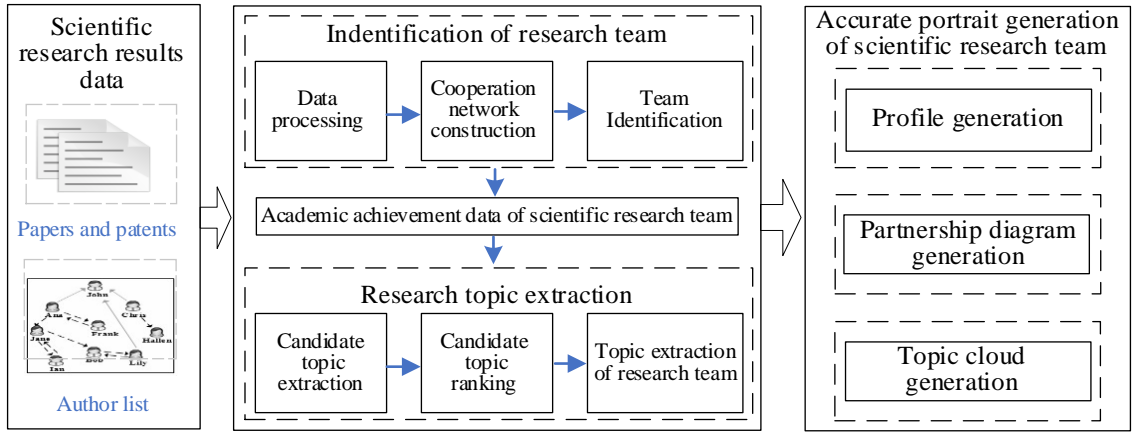

Fig. 1 Framework of accurate stereo portrait algorithm for scientific research team

Collect scientific research achievements (including papers and patents) and author lists of scientific and technological achievements in various fields to form a scientific and technological achievements data set. For the scientific and technological achievements data set, the iterative intermediate centrality method is used to identify the leader of the scientific research team, the core personnel and non-core personnel of the scientific research team. The information of the identified scientific research team is statistically analyzed to form a team profile. According to the list of authors of the scientific research team's scientific research results, the co-occurrence information of authors, the authors and supervisors in the master's and doctoral dissertations of scientific research team members, and the cooperation information and mentoring relationship between the leaders of the scientific research team and team members are extracted. Combing with topic extraction method TF-IDF and the graph-based topic extraction method TextRank, comprehensively consider the statistical and semantic features of the words to score topics and extract the topics from the scientific research team's achievement data set, and generate the research topic of the scientific research team. Finally, the accurate stereo portrait of the scientific research team is constructed.

### 3.1 Identification of the scientific research team

The team leader is the core and soul of the scientific research team, and the identification of the leader of the scientific research team is a crucial step in the identification of the scientific research team. This paper identifies the leader of the scientific research team based on the iterative intermediate centrality ranking method, and identifies the scientific research team members through the 2-faction and snowball method, so as to realize the identification and mining of the scientific research team.

The scientific research team identification module uses the list of scholars co-authored as input to construct a scientific research cooperation network, calculate the intermediate centrality ranking of all nodes in the scientific research cooperation network, select the leader of the research team, and use the 2-faction method to identify the core connected to the team leader. Based on scientific researchers, through the snowball method, identify non-core scientific researchers based on the leader of the scientific research team and core scientific researchers, and form a scientific research team.

The specific description of the identification algorithm of the scientific research team is shown in Table 1.

Table 1 Research team identification process

| Process of scientific research team identification module |
|---|
| **Input**：List of scholarly co-authored works |
| **Output**：Scientific research team |
| 1 Get a list of co-authors of scientific research results and build a co-authorship network； |
| 2 In a co-authorship network, compute the betweenness centrality rank of all nodes； |
| 3 Record the scientific researcher whose median centrality value is top1 as the leader of the scientific research team in the institution, and then delete the node in the overall scientific research cooperation network； |
| 4 Continue to calculate the intermediate centrality ranking of the remaining nodes, and determine whether the intermediate centrality value of the top1 node is not greater than 1. If yes, go to step 4; if not, go to step 2 again； |
| 5 Output research team leader； |
| 6 Use the 2-Faction approach to identify the core research member connected to the research team leader； |
| 7 Set up a scientific research cooperation network with a co-authoring frequency of 2, and use the snowball method to scroll down one layer from the research team leader and core researchers to identify non-core researchers. |

### 3.2 Research topic extraction of research team

The research topic of the scientific research team reflects the research direction of the entire scientific research team and is an important part of the scientific research team. The representative topics in the scientific research team are extracted to describe the scientific research team. Most of the traditional topic extraction methods only rely on the limited word frequency, document frequency and other statistical information of candidate topics, and do not consider the use of candidate topics in the corresponding fields in the scientific and technological achievements data, so the accuracy of topic extraction is not accurate enough. This paper combines the statistical-based TF-IDF topic extraction method with the graph-based TextRank method to extract topic words from the research team's results. The framework of the research topic extraction module of the scientific research team is shown in Figure 2.

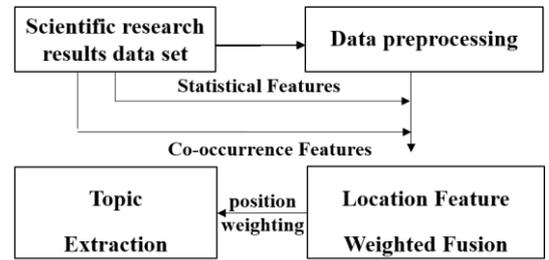

Fig 2. Research topic extraction framework for research team

(1) Topic extraction based on statistical features

TF-IDF weights all candidate topics in the text, and sorts the topics according to the weights. Assuming D is a set of documents, data preprocessing operations such as word segmentation, part-of-speech tagging and removal of stop words are performed on the given document $d$. Retain nouns, verbs, adjectives, gerunds, adverbs and other parts of speech, and finally get $n$ candidate topics, calculate the word frequency $tf_{w_i}$ of each keyword $w_i$ in document $d$ and inverse document frequency $idf_{w_i}$. The calculation formulas of $tf_{w_i}$ and $idf_{w_i}$ are as follows:

$$tf_{w_i} = d_{w_i} / |d| \qquad (1)$$

$$idf_{w_i} = \log(\frac{D}{D_{w_i}+1}) \qquad (2)$$

Where, $d_{w_i}$ is the number the word appears in document $d$, $|d|$ is the total number of word of the document $d$. D is the total number of documents in the document collection. $D_{w_i}$ is the number of documents in which the word $w_i$ appears in the document collection.

Calculate the $tfidf(w_i)$ values of all words in document d, and the calculation formula is as follows:

$$tfidf(w_i) = tf_{w_i} * idf_{w_i} \qquad (3)$$

Repeat the above steps to obtain the $tfidf$ values of all topics. Finally, the calculation results of candidate topics are sorted in reverse order, and the top N topics are obtained as text topics.

(2) Topic extraction based on graph features

TextRank is a graph-based topic word extraction method. It uses the words in the document as nodes to construct a graph, and uses the co-occurrence information between the words in the document. If two words co-occur, there will be an edge in the graph.

$$tr(w_i) = (1 - d) + d * \sum_{j \in In(V_i)} \frac{w_{ji}}{\sum_{V_k \in Out(V_j)} w_{jk}} tr(w_j) \quad (4)$$

Where, $tr(w_i)$ is score of word $w_i$, $d$ is damping coefficient. Take 0.85 in the text.

Given a document, first word segmentation and part-of-speech tagging are performed, and a single word as a node is added to the graph. Second, set a grammar filter to add nouns, verbs, adjectives, gerunds, and adverbs of part-of-speech to the graph. Third, words appear in a window form an edge with each other. Finally, based on formula (4), through a certain number of iterations, different word nodes will have different weights in the end, and words with high weights are used as topics.

(3) Topic Extraction based on Feature Position Weighting

The topic extraction result of TF-IDF and the topic extraction result of TextRank are taken as the final extracted topic set, and the position-weighted summation of each feature is performed using formula (5) to obtain the final score of each keyword.

$$S(w_i) = \frac{1}{pos_{tfidf}(wi)} * tfidf(wi) + \frac{1}{pos_{tr}(wi)} * tr(wi) \quad (5)$$

**3.3 The scientific research team's accurate stereo portrait generation algorithm**

The accurate stereo portrait of the scientific research team consists of three parts: the introduction of the scientific research team, the construction of the team cooperation relationship diagram, and the research topic of the scientific research team. This paper designs a template for each part. The research team template mainly includes the total number of research teams, the main research fields of the research team, the number of projects hosted by the team leader, the total number of team projects, the total number of published papers, the number of citations, and the total number of patents. The number of conference papers, journal papers, and monographs published by the team; the team cooperation relationship diagram template includes the cooperative relationship between the team leader and the research team and the mentoring relationship between them, the institution and discipline of the team members; research team topics. The template mainly displays the research topics of the scientific research team in the form of topic word clouds.

The scientific research team identification module identifies the basic information of the leader of the research team and members of the research team, including the name of the research team, affiliation, research direction, fund information of the team leader and team members, scientific research achievements (papers, patents, master's degree etc.), citation information, and monograph information.

Statistically analyze the basic information of the scientific research team to generate a brief introduction of the scientific research team, and mine the cooperative relationship and mentoring relationship between the scientific research team leader and members through the master and doctoral dissertation information of the team leader and team members; extract the scientific research team based on the research topic of the research team, generate the research topic word cloud map through the research topic of the scientific research team. Combine scientific research team introduction, scientific research team cooperation relationship diagram and scientific research team research topic words to generate accurate stereo portrait of scientific research team.

**4 Experimental results**

**4.1 Effectiveness analysis of scientific research team identification**

(1) Data preprocessing

We collected the achievement data (including papers and patents) of 2000 projects on the website of the National Natural Science Foundation in China(in which there is a situation where one project leader presides over multiple projects) and the author information list of each achievement. There are a total of 172,879 records, and each record includes such information as the title of the paper, the title of the patent, the author, and the author's unit. If the obtained data has invalid data, such as no-author or single-author paper data, such data will be deleted.

The premise of the identification of the scientific research team is to disambiguate the author data, and the author disambiguation needs to be combined with the institutional data to clean the institutional data and clean the author data. The author information in the obtained data includes Chinese authors and English authors. For English data, the author's name and institution are aligned with the Chinese data to convert it into a unified Chinese author to facilitate subsequent processing. After processing the original data, 146,868 records were obtained for this study, with a total of 102,312 authors.

(2) Identification of the scientific research team

Use the preprocessed data to construct an overall scientific research co-authorship network. From the overall network, extract the co-authoring network with the minimum threshold of author's publication volume of 10 and the minimum threshold of author's co-authoring frequency of 5 as the initial scientific research co-authoring network, calculate the intermediate centrality ranking of all nodes, and select the scholar with the highest ranking as scientific research. Remove the scholar from the co-authorship network, continue to calculate the median centrality ranking, and select the leader of the research team with the highest ranking.

**Table 2 Team leader identification results**

| Ranking | Team leader | centrality | In line with objective facts or not |
|---|---|---|---|
| 1 | Team one's leader | 478 | yes |
| 2 | Team two's leader | 134 | yes |
| 3 | Team three's leader | 90 | yes |
| 4 | Team four's leader | 72 | yes |
| 5 | Team five's leader | 58 | yes |

The iterative-based intermediate centrality ranking method selects the team leader through repeated iterations, calculates the intermediate centrality value ranking by looping, and only extracts the highest-ranked node as the team leader each time, then removes the node from the network and continues the iterative calculation, until the intermediate centrality value of the remaining nodes is not greater than 1, and a better identification result of the team leader is obtained.

In the initial co-authorship network, a set of nodes that have a co-authorship relationship where the team leader is selected, and the results of the scientific research team leader identified by the iterative intermediate centrality method are used as the base point. And the 2-faction method is used in the overall scientific research cooperation network. Identify the core scientific researchers of the team, and verify the background information of the identified core scientific researchers, such as the unit, number of cooperation, and cooperation relationship. From the identification results of the core researchers of each research team, it can be seen that the core researchers of each research team are mainly research colleagues from the same institution as the team leader, supplemented by cross-institutional researchers. In the overall co-authorship network, with the team leader of the scientific research team and each author of the core member as the apex, scroll down one layer to get the non-core members of the scientific research team.

**4.2 Effectiveness analysis for research topic extraction of scientific research team**

(1) Dataset introduction

The project names and abstracts of 200 general projects and 300 youth science fund projects were selected as the source documents extracted from candidate subject headings to form Data Set 1 and Data Set 2. The description of the data sets is shown in Table 3.

**Table 3. Dataset description**

| Dataset | Dataset1 | Dataset2 |
|---|---|---|
| information | name,abstract,keyword | name, abstract,keyword |
| number | 200 | 300 |
| project type | surface | youth |

(2) Evaluation indicators

In order to evaluate the effectiveness of the scientific research team topic extraction algorithm, the accuracy rate, recall rate, and F1 value are selected as evaluation indicators, and the calculation formula is as follows:

$$Precision = \left(\frac{TP}{TP+FP}\right) \tag{6}$$

$$Recall = \left(\frac{TP}{TP+FN}\right) \tag{7}$$

$$F1 = \left(\frac{2*Precision*Recall}{Precision+Recall}\right) \tag{8}$$

Where, TP is the number of correct topics in the extracted topics, FP is the number of incorrect topics in the extracted topics; FN is the number of correct topics that are misjudged.

(3) Contrast algorithm

The method in this paper combines the TF-IDF and TextRank methods, abbreviated as TF-TR, and the TF-IDF algorithm and the TextRank algorithm are selected as the comparison algorithms for verification.

TF-IDF: Consider the statistics of words, calculate the score based on the frequency of the word in the document and the frequency of document.

TextRank: A graph-based topic extraction method, TextRank uses the co-occurrence information between words in a document to model the relationship between words.

(4) Experimental results and analysis

The Precision@n (abbreviated as P@n), Recall@n (abbreviated as R@n) and F1@n are used as the evaluation criteria for the topic extraction results, and $n$ is the number of extracted topic words. In the text, $n$ takes 1, 3, 5, and 10. Invite graduate students in related fields to participate in the evaluation of the results of the topic extraction of the scientific research team. In the two scientific research project data sets, the module verification results of the accuracy, recall, and F1 of the project research topic extraction are shown in Table 4-7 and Figure 3-4.

Table 4. Validation results of research topic precision on dataset 1

|  | P@1 | P@3 | P@5 | P@10 |
|---|---|---|---|---|
| TF-IDF | 0.7550 | 0.6517 | 0.5580 | 0.4025 |
| TextRank | 0.73 | 0.63 | 0.5170 | 0.3690 |
| TF-TR | 0.7782 | 0.6837 | 0.5792 | 0.4203 |

Table 5. Validation results of research topic recall on dataset 1

|  | R@1 | R@3 | R@5 | R@10 |
|---|---|---|---|---|
| TF-IDF | 0.0830 | 0.2119 | 0.3018 | 0.4328 |
| TextRank | 0.0800 | 0.2067 | 0.2804 | 0.3959 |
| TF-TR | 0.1011 | 0.2432 | 0.3327 | 0.4479 |

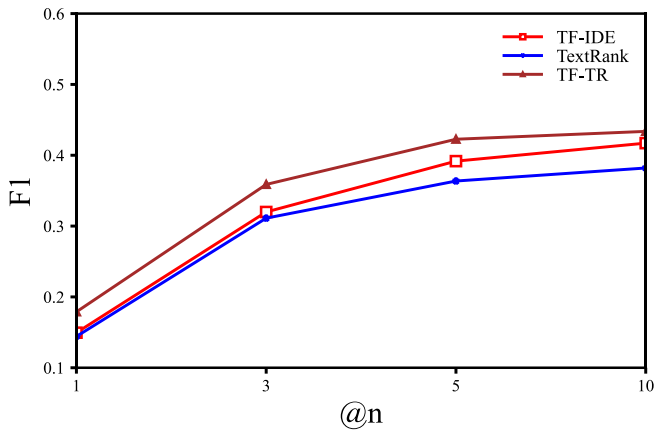

Fig 3. Validation results of research topic F1 on dataset 1

Table 4 shows the accuracy of research topic extraction for scientific research projects on dataset 1, and Table 5 shows the recall for research topic extraction for scientific research projects on dataset 1. From the results in Table 4-5, we can see that our comprehensive topic extraction method TF-IR which considers the co-occurrence information between statistical information and words, which further performs position weighted summation on topics, has better topic extraction effect than TF-IDF algorithm and TextRank algorithm.

Figure 3 shows the F1 extracted from the research topics of scientific research projects on Data Set 1. The F1 combines the accuracy and the recall to evaluate the extracted topics. It can be seen from Figure 3 that TF-IR topic extraction is effective.

Table 6. Validation results of research topic precision on dataset 2

|  | P@1 | P@3 | P@5 | P@10 |
|---|---|---|---|---|
| TF-IDF | 0.75 | 0.6644 | 0.5533 | 0.4040 |
| TextRank | 0.7167 | 0.6022 | 0.5147 | 0.3670 |
| TF-TR | 0.7783 | 0.6832 | 0.5812 | 0.4351 |

Table 7. Validation results of research topic recall on dataset 2

|  | R@1 | R@3 | R@5 | R@10 |
|---|---|---|---|---|
| TF-IDF | 0.078 | 0.2108 | 0.2915 | 0.4200 |
| TextRank | 0.0754 | 0.1888 | 0.2680 | 0.3810 |
| TF-TR | 0.0920 | 0.2315 | 0.3045 | 0.4431 |

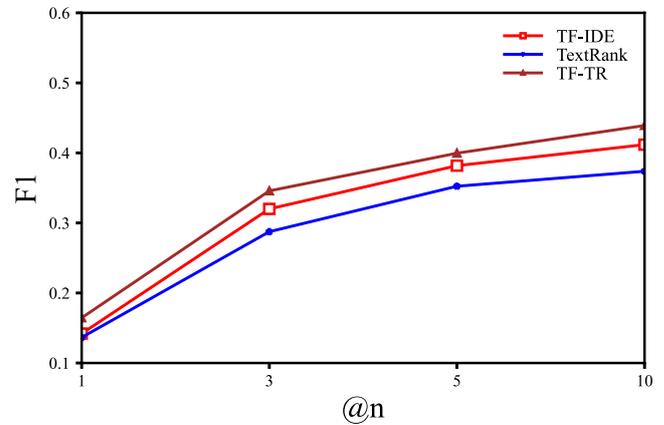

Fig 4. Validation results of research topic F1 on dataset 2

Table 6-7 shows the accuracy and recall of research topics extracted from scientific research projects on Data Set 2, and Figure 4 shows the F1 of research topics extracted from scientific research projects on Data Set 2. Table 6-7 and Figure 4 results the topic extraction performance on dataset 1 shows the same effect, that is, TF-IR has better topic extraction effect than TF-IDF and TextRank. This shows the effectiveness of the topic extraction method TF-IR on the research team achievement dataset.

### 4.3 Accurate stereo portrait generation of the scientific research team

Based on the identified scientific research team, obtain all the information of the team, and generate a team portrait including the scientific research team introduction, the scientific research team cooperation relationship diagram, and the research topics of the research team.

Through statistical analysis of the basic information of the scientific research team, a brief introduction of the scientific research team is formed. The team introduction shows the total number of team members, the research field of the team, the number of projects hosted by the team leader, the total number of team projects, the total number of published papers, the number of citations, and the total number of patents. The number of conference papers, journal papers, and monographs published by the team shows the basic information of the team. Build a research team cooperation relationship diagram to show the cooperation relationship between the team leader and the research team, the mentoring relationship between them, and the institutions and disciplines to which the team members belong. Based on the topic extraction module of the scientific research team, the research topics are extracted from the scientific research results of the scientific research team to form a topic word cloud, which depicts the main research topics of the scientific research team.

## 5 Conclusion

In response to the practical needs of scientific research management departments to find excellent scientific research teams due to the establishment of scientific research projects, accurate stereo portrait generation algorithm of scientific research teams is proposed. The templates of the introduction of the research team, the cooperation relationship of the research team, and the research theme of the research team are designed to intuitively describe the research team. The scientific research team identification module accurately identifies the scientific research team based on the massive scientific research results resources, performs statistical analysis on the scientific research team information, generates the scientific research team profile and the scientific research team cooperation relationship diagram based on the template. The research topic extraction module of the research team, and combines the characteristics of the scientific research data. Considering the statistical characteristics of words and the correlation between words, the research topic of the research team is extracted based on the scientific research achievement data of the scientific research team, and the word cloud of the research topic is generated based on the template, and an accurate three-dimensional portrait of the scientific research team is generated. Based on the proposed accurate three-dimensional portrait algorithm of scientific research teams, the scientific research team is identified on the real scientific research results, and the scientific research team is constructed and visualized. The experimental results verify the effectiveness of the algorithm.